\documentclass[aps,prl,showpacs,twocolumn]{revtex4}
\usepackage{epsfig,dsfont,amssymb,amsmath,amsthm,amsfonts,amsbsy,mathrsfs}
\usepackage{graphicx}
\usepackage{color}
\usepackage{bm}

\begin{document}
\title{Structural Signature of Slow and Heterogeneous Dynamics in Glass-Forming Liquids}
\author{Yan-Wei Li}
\affiliation{State Key Laboratory of Polymer Physics and Chemistry, Changchun Institute of Applied Chemistry, Chinese Academy of Sciences, Changchun 130022, China}
\author{You-Liang Zhu}
\affiliation{State Key Laboratory of Polymer Physics and Chemistry, Changchun Institute of Applied Chemistry, Chinese Academy of Sciences, Changchun 130022, China}
\author{Zhao-Yan Sun}
\email{zysun@ciac.ac.cn}
\affiliation{State Key Laboratory of Polymer Physics and Chemistry, Changchun Institute of Applied Chemistry, Chinese Academy of Sciences, Changchun 130022, China}

\date{\today}

\begin{abstract}
One of the central problems of the liquid-glass transition is whether there is a structural signature that can qualitatively distinguish different dynamic behaviors at different degrees of supercooling. Here, we propose a novel structural characterization based on the spatial correlation of local density and we show the locally dense-packed structural environment has a direct link with the slow dynamics as well as dynamic heterogeneity in glass-formers. We find that particles with large local density relax slowly and the size of cluster formed by the dense-packed particles increases with decreasing the temperature. Moreover, the extracted static length scale shows clear correlation with the relaxation time at different degrees of supercooling. This suggests that the temporarily but continuously formed locally dense-packed structural environment may be the structural origin of slow dynamics and dynamic heterogeneity of the glass-forming liquids.
\end{abstract}

\pacs{61.20.Ja, 61.20.Lc, 64.70.P-}

\maketitle

Upon cooling, glass forming liquids display markedly slow and heterogeneous dynamics but unperceivable change in its static pair correlations~\cite{Berthier0,Binderbook}. Although intense attention has been paid to explore the structural origin of slow dynamics and dynamical heterogeneity, a more convincing view is still lack and the relationship between structure and dynamics seems to be quite complex~\cite{Barrat,Karmakar0}. It is still unclear whether the glass transition is due to an underlying thermodynamic phase transition or purely dynamic process in nature. The related mode-coupling theory, which predicts the dynamics solely based on the static pair correlations, is found to be failed to explain the nonperturbative effect of attractive forces in supercooled liquids, suggesting the inefficiency of pair correlation when describing the structural change of glass-forming liquid~\cite{Berthier1,Berthier2,Berthier3}.

To unveil static quantities, several different ideas have been proposed. The locally preferred structural order, accompanied with little density change, is claimed to be the cause of the slow dynamics, as growing in its length scales with increasing the degree of supercooling in several systems~\cite{Tanaka0,Tanaka1,Tanaka2,Coslovich0}. However, this method shows clearly system dependences~\cite{Hocky1}, and is hard to be extended to different glass-formers (e.g., the high dimensional liquids~\cite{Bruning} and the strong frustrated glass formers~\cite{Charbonneau}). Recently, a more general method investigating the point-to-set (PTS) correlations in the amorphous system emerged and attracted lots of attention~\cite{Biroli0,Biroli1,Kob1,Hocky,Tarjus1,Li}. It is worth noting that the obtained modest increase of the static PTS length scale ($\xi_{PTS}$) shows clear decoupling with the dramatic increase of the dynamical length characterizing spatial heterogeneities of the glass-formers~\cite{Tarjus1, Li,Tanakaadd}, which arouses heated debates on the relationship between structure and dynamics of glass-formers. Clearly, exploring more general structural signatures of slow dynamics is crucial to understand the structural origin of the slow dynamics and dynamic heterogeneity for glass-forming liquids.

For glass formers, one of the most pronounced features is that particles are trapped in the transient cages~\cite{cage1,cage2,cage3,cage4}, which leads to the rigid nature of amorphous materials. This brings an intriguing question that whether this localization dynamics corresponds to the change of the local structural environment. A perfect candidate, i. e., the local density ($\rho_{l}$), which can be determined from the Voronoi cell volume ($v$), can provide such local information for each particle. Previous investigations found weak correlation with dynamics based on per-particle local geometry~\cite{Bernini1,Bernini2}. Up to now, neither indications of the spatial density correlation nor the link between the local structure characterized by $\rho_{l}$ and glassy dynamics was reported.

In this work, instead of concentrating on the first coordination shell local geometry, we study the spatial correlation of the relatively dense-packed structural environment and explore its links with the slow dynamics as well as dynamic heterogeneity in glass-forming liquids by means of molecular dynamics simulation. We achieve positive results of the above relationship by studying three dimensional (3D) Kob-Andersen binary Lennard-Jones mixture (LJ) ~\cite{Kob2} and its Weeks-Chandler-Andersen (WCA) truncation~\cite{WCA} over the density range from $\rho=1.2$ to $\rho=1.9$ (see system details in the Supplemental Material~\cite{SM}). These two systems have different dynamics at low densities in the supercooled regime while this difference shrinks as increasing $\rho$ and almost disappears at $\rho=1.9$~\cite{Berthier1,Berthier2,xuning}. On the other hand, in the whole studied density range, the two systems have nearly identical two body static correlations at each temperature. Thus these two systems provide a perfect testbed to study whether the local density is sensitive to structural differences and whether there exists a direct link between structure and dynamics for glass-formers. We demonstrate that LJ system has a larger size of clusters ($<N_{c}>$) formed by the relatively dense-packed particles than WCA system at lower density. This corresponds to the slower dynamics of the LJ system. Moreover, the difference of the value of $<N_{c}>$ for these two systems shrinks synchronously with the dynamics as increasing $\rho$. The extracted static length scale is found to correlate well with the degree of supercooling and also the relaxation time in the studied temperature and density range, suggesting that the locally dense-packed environment might be the main reason of slow dynamics and dynamical heterogeneity in glass-formers.

\begin{figure}[tb]
 \centering
 \includegraphics[angle=0,width=0.46\textwidth]{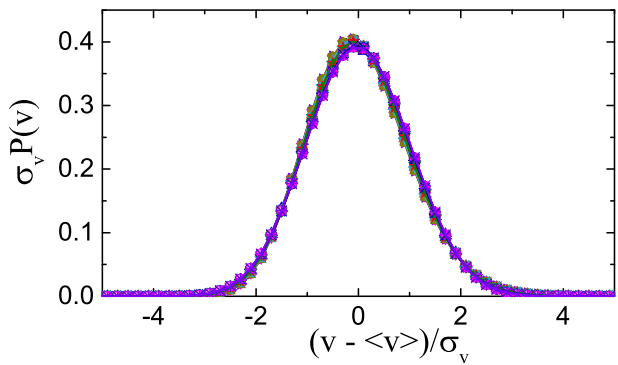}
 \caption{(color online). Scaled distribution of Voronoi cell volumes, $\sigma_{v}P(v)$ as a function of the normalized  Voronoi volume, ($(v-<v>)/\sigma_{v}$) at $30$ different state points of variable $T$ and $\rho$ (ranging from high $T$ liquid to low $T$ supercooled states for both LJ and WCA systems). All the results are for $A$ particles.}
\end{figure}

To find which particles are densely packed in the system, we first implement the Voronoi constructions~\cite{voronoi} and calculate the Voronoi cell volume $v$ for each particle. Follow the protocol of Ref.~\cite{Starr0}, we investigate the statistical properties of the Voronoi cells by calculating the scaled distribution of $v$ (see Fig. 1). Here, $v$ is shifted by the average Voronoi volume $<v>$ and then normalized by the standard deviation $\sigma_{v}$ ($\sigma_{v}^{2}=<v^{2}>-<v>^{2}$). As expected, the distributions ($P(v)$) only show little difference for various T and $\rho$ and are not sensitive to the system difference, consistent with Refs.~\cite{Starr0,Cheng}. Such kind of scaling collapse behavior is expected to be universal in dense liquids with relatively harder interactions, and deviations have been found in soft-potential colloidal suspensions~\cite{Starr1}. Our results indicate similar intrinsic amorphous structure from the statistical view at different $T$ and $\rho$ of the LJ and WCA glass formers. Although neither the static pair correlations nor the statistical distribution of the Voronoi volumes accounts for the significant dynamic differences of these two systems, other subtle characterizations such as the higher order structural correlations~\cite{Coslovich0,xu}, the PTS correlation length~\cite{Hocky} and properties of zero-temperature glasses~\cite{xuning} can also capture the structural differences.

The distributions ($P(v)$) show little dependence on $T$ and $\rho$ for both LJ and WCA systems, however, the growth of the static correlations of the relatively dense-packed particles exhibit marvelous connection with slow dynamics upon cooling. First, we emphasize that the local density is not correlated with local ordering, as has been proved in Refs.~\cite{Tanaka0,Russo,Tanpeng}. We define $13\%$ of particles with the largest local density $\rho_{l}$ ($\rho_{l}=1/v$) as the relatively dense-packed particles (The qualitative results are not sensitive to the choice of the threshold value $13\%$). We also use other criterions to define the densely packed particles (e. g., particles with Voronoi volume satisfying $(v-<v>)/\sigma_{v}<-1.2$ are defined as the densely packed ones) and we can obtain the very similar results (see the Supplemental Material~\cite{SM} for details). For type A and type B particles, the range of $\rho_{l}$ is distinct, thus we sort type A and type B particles separately and choose $13\%$ of type A particles and $13\%$ of type B particles with the largest local density respectively as the densely packed particles. Two densely packed particles are considered as belonging to the same cluster if they are neighbors defined by the Voronoi method. The number of densely packed particles that are belonging to a cluster is defined as the single cluster size $N_{c}$. The number averaged cluster size $<N_{c}>$ is defined as
\begin{equation}
<N_{c}>=\frac{\sum_{N_{c}=1}^{\infty}N_{c}P(N_{c})}{\sum_{N_{c}=1}^{\infty}P(N_{c})},
\end{equation}
where $P(N_{c})$ is the probability of finding a single cluster size $N_{c}$, and $\sum_{N_{c}=1}^{\infty}P(N_{c}) = 1$. The value of $<N_{c}>$ can reflect the degree of spatial correlations between dense-packed particles.

In Fig. 2, we show the temperature dependence of the relaxation time  $\tau_{\alpha}$ and that of the number averaged cluster size $<N_{c}>$ at $\rho=1.2$ and $\rho=1.9$ for both LJ and WCA systems. $\tau_{\alpha}$ is defined as $F_{s}(q_{p},t=\tau_{\alpha})=1/e$ (where $F_{s}(q_{p},t)$ is the intermediate scattering function of $A$ particles as shown in the Supplemental Material~\cite{SM}). It is clear that both $\tau_{\alpha}$ and $<N_{c}>$ increase as the temperature is lowered, implying that $<N_{c}>$ can reveal unambiguous structure difference between the supercooled liquid and normal liquid as an effective static signature. At $\rho=1.2$, LJ system has larger $\tau_{\alpha}$ and thus its dynamics is slower than the WCA system at the same $T$ in the supercooled states (see Fig. 2(a)). Correspondingly, $<N_{c}>$ is larger for LJ system (see Fig. 2(c)). On the other hand, the $<N_{c}>$ difference of the two systems is small but far from negligible at high $T$, although $\tau_{\alpha}$ of the two systems is quite close. This is quite reminiscent of the predictions from the Arrhenius law that the LJ and WCA system have distinct dynamics in the high $T$ regime at $\rho=1.2$, since the active energy of the two systems is different~\cite{Berthier2}. At $\rho=1.9$, both $\tau_{\alpha}$ and $<N_{c}>$ are very close for the two systems, implying both dynamics and its structural signature are similar (see Fig. 2(b) and 2(d)). Our results show direct evidence of the correlation between static property characterized by the locally dense-packed structural environment and dynamic behavior at different degrees of supercooling in a large density range. It should be noted that, different from the static PTS correlation length~\cite{Biroli1,Kob1}, the present $<N_{c}>$ data undergoes no fitting or further processing, and this unambiguously avoids some sorts of artifacts in the interpretation of the data. This local density based spatial correlations can be an overwhelming tool to characterize the static nature of glass-formers in some systems such as the high dimensional liquids~\cite{Bruning} or strong frustrated glass-formers~\cite{Charbonneau}, in which the bond-orientational order is not obvious.

\begin{figure}[tb]
 \centering
 \includegraphics[angle=0,width=0.47\textwidth]{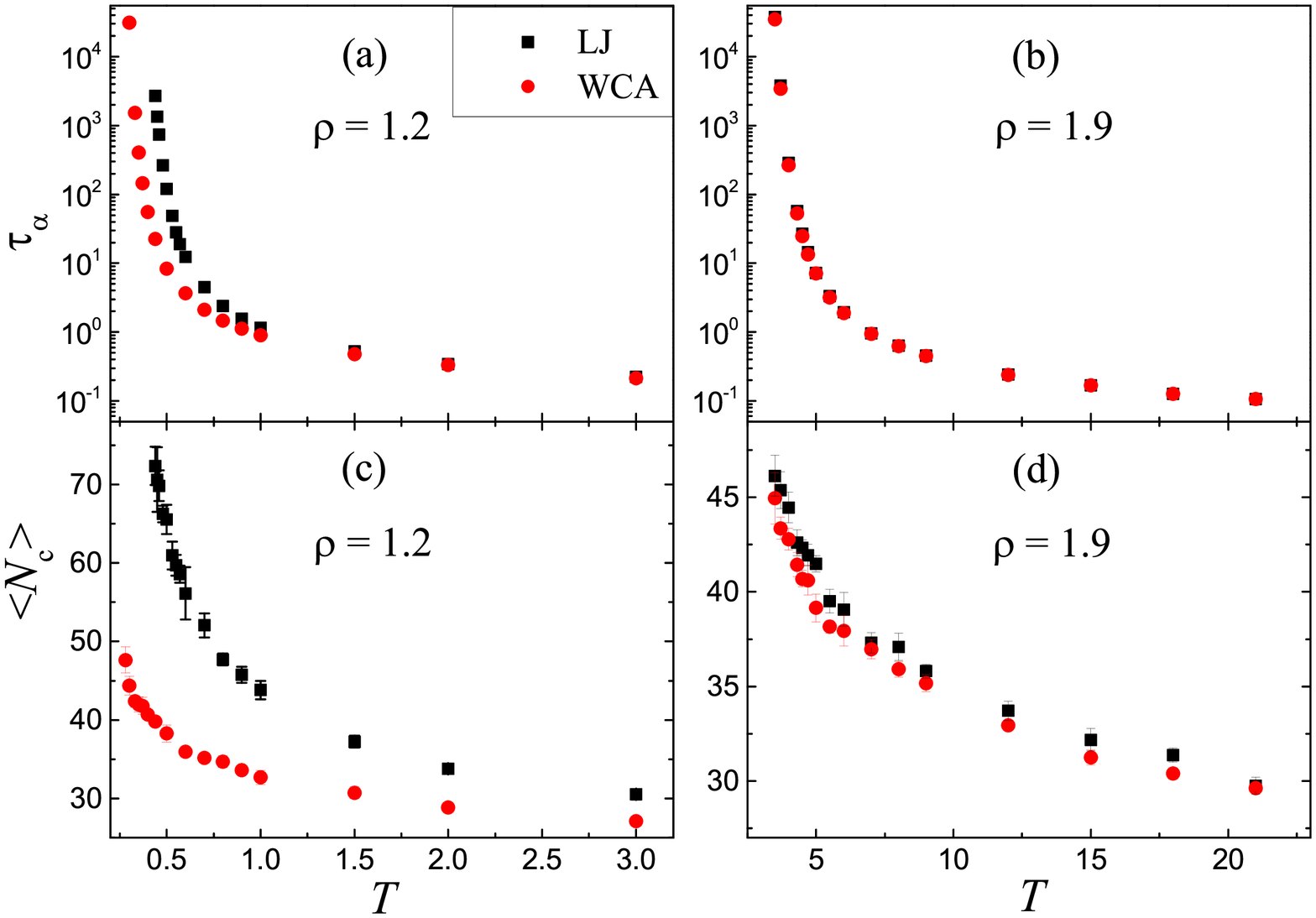}
 \caption{(color online). $T$ dependence of the relaxation time $\tau_{\alpha}$ of the LJ and WCA systems for $\rho=1.2$ (a) and $\rho=1.9$ (b), and $T$ dependence of the number averaged cluster size $<N_{c}>$ of the LJ and WCA systems for $\rho=1.2$ (c) and $\rho=1.9$ (d). The black squares and red circles are for LJ and WCA systems, respectively. $<N_{c}>$ is obtained by averaging over 1000 configurations in each run, and the error bars in (c) and (d) are obtained by averaging over 8 independent runs.}
\end{figure}

\begin{figure}[tb]
 \centering
 \includegraphics[angle=0,width=0.40\textwidth]{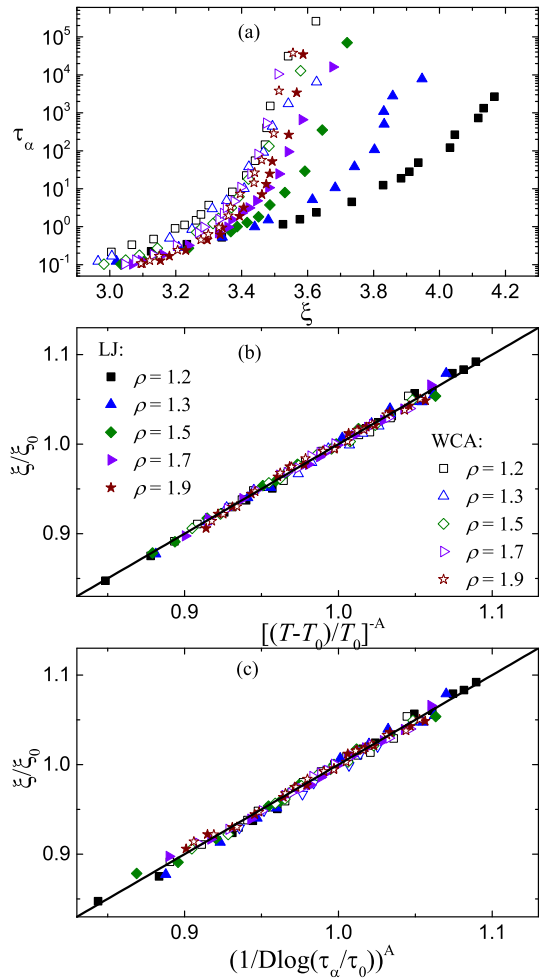}
 \caption{(color online). (a) The $\alpha$ relaxation time $\tau_{\alpha}$ vs the static length scale $\xi$ for all the density range studied for both the LJ and WCA systems. (b) The rescaled length scale $\xi/\xi_{0}$ versus $[T_{0}/(T-T_{0})]^{A}$. The black solid line is the relation of $\xi/\xi_{0}=[T_{0}/(T-T_{0})]^{A}$. (c) Relation between $\xi/\xi_{0}$ and $[1/Dlog(\tau_{\alpha}/\tau_{0})]^{A}$. The black solid line is the relation of $\xi/\xi_{0}=[1/Dlog(\tau_{\alpha}/\tau_{\alpha})]^{A}$. Legends in (a), (c) are the same as those in (b).}
\end{figure}

We define the static correlation length of locally dense structural environment as $\xi=\sqrt[3]{<N_{c}>}$ based on our three dimensional glass model. Similar definition in two dimensional glass-forming liquids has been used in defining the length scale of the hexatic structural order~\cite{Tanaka0,Tanaka1,Tanaka2}. Next, we study the growth of $\xi$ as increasing the degree of supercooling and its link to slow dynamics for all the density range we studied in both LJ and WCA systems. Fig. 3(a) illustrates the behavior of $\tau_{\alpha}$ as a function of the static length scale $\xi$. It is clear that $\tau_{\alpha}$ grows monotonically with $\xi$ in the whole temperature range for all the systems, which is consistent with the behavior of other reported static length scales~\cite{Biroli1,Kob1,Hocky,Tarjus1,Li,Tanaka0,Tanaka1,Tanaka2,Tanakaadd} as increasing the degree of supercooling. Moreover, the data in Fig. 3(a) shows that the difference between LJ and WCA systems is pronounced at $\rho=1.2$, but shrinks as increasing $\rho$, indicating that attractive tails have a remarkable impact on the system at relatively low density, similar to the related reports~\cite{Berthier1,Berthier2}.
Inspired by the temperature dependence of the length scale of the structural order~\cite{Tanaka0}, we expect that $\xi$ behaves as
\begin{equation}
\xi=\xi_{0}[T_{0}/(T-T_{0})]^{A},
\end{equation}
where $\xi_{0}$ and $A$ are the fitting parameters. Both the value of them reflect the intrinsic property of the system and are quantitatively impacted by the definition of dense-packed particles. Detailed discussions on $\xi_{0}$ and $A$ can be found in the Supplemental Material~\cite{SM}. $T_{0}$ is the critical glass transition point fitted by Vogel-Fulcher-Tamman (VFT) relation
\begin{equation}
\tau_{\alpha}=\tau_{0}exp[DT_{0}/(T-T_{0})],
\end{equation}
where $\tau_{0}$, $D$ and $T_{0}$ are the fitting parameters. $D$ is the fragility parameter, and smaller value of $D$ corresponds to a more fragile glass former. All our simulation data can be well fitted by VFT equation (see Fig. S2 of the Supplemental Material~\cite{SM}). Fig. 3(b) plots the relation between static length scale and temperature of the LJ and WCA systems in the whole density range we studied. Interestingly, all curves collapse when we plot $\xi/\xi_{0}$ against $[T_{0}/(T-T_{0})]^{A}$, implying a direct link of the static length scale $\xi$ and the degree of supercooling. Combining Eq. (2) and Eq. (3), we obtain the scaling relation of the static length scale $\xi$ and the relaxation time $\tau_{\alpha}$
\begin{equation}
\xi=\xi_{0}[1/Dlog(\tau_{\alpha}/\tau_{0})]^{A},
\end{equation}
The results of this relation at different densities in the LJ and WCA systems are shown in Fig. 3(c). Clearly, the $\xi - \tau_{\alpha}$ relation supports the scaling argument of Eq. (4). As a result, our findings provide useful clues to the underlying causes of the slow dynamics in the supercooled regime. The static length scale of the densely structural environment plays a key role in determining the dynamics of the glass-forming liquids. Moreover, we test our results in other two glass models and we can also find similar collapses in the relations of both $\xi - T$ and $\xi - \tau_{\alpha}$ (see the last section of the Supplemental Material~\cite{SM}). To our knowledge, this is the first finding of the link between density based spatial correlations and dynamics.

\begin{figure}[tb]
 \centering
 \includegraphics[angle=0,width=0.45\textwidth]{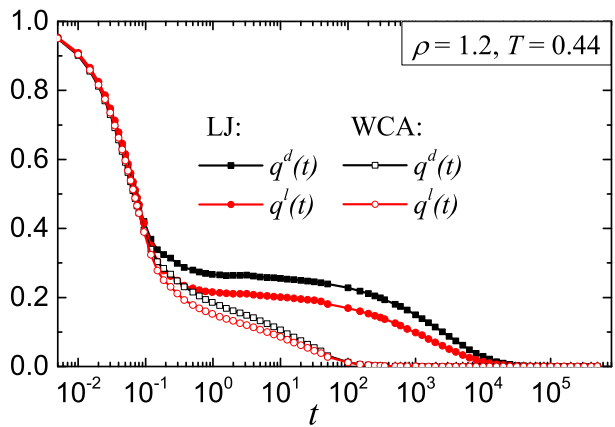}
 \caption{(color online). Time dependence of the self part of dense-packed particles overlap $q^{d}(t)$ and that of loose-packed particles overlap $q^{l}(t)$ at $T=0.44$ and $\rho=1.2$ for both LJ and WCA systems.}
\end{figure}

Another crucial question is the relation between the locally structural environment and the dynamic heterogeneity. To characterize the dynamics of the dense-packed and loose-packed particles, the self part of dense-packed particles overlap $q^{d}(t)$ and loose-packed particles overlap $q^{l}(t)$ are defined as follows. We partition the simulation box into small cubic boxes with side length $l_{s} = 0.36$, so that there is no more than one particle in each small cubic box. Then we define a binary digit $n_{i}^{\alpha}(t)$ to specify whether the same dense-packed particle or loose-packed particle occupies the cell $i$ at times $0$ and $t$.
\begin{equation}
q^{\alpha}(t)=\frac{\sum_{i}<n_{i}^{\alpha}(t)n_{i}^{\alpha}(0)>}{\sum_{i}<n_{i}^{\alpha}(0)>},
\end{equation}
where the sum runs over all the small cubic boxes in the system. The superscript $\alpha = d$ is for dense-packed particles and $\alpha = l$ is for loose-packed particles, respectively. Here, we also define $13\%$ of particles with the largest local density as the dense-packed particles and $13\%$ of particles with the smallest local density as loose-packed particles.

In Fig. 4, we show the time dependence of $q^{d}(t)$ and $q^{l}(t)$ at $T=0.44$ and $\rho=1.2$ for both LJ and WCA systems. Similar to the self-intermediate scattering function as shown in Fig. S1~\cite{SM}, the overlaps show clear two-step decay for both LJ and WCA systems, indicating that both two systems are in the supercooled regime. Moreover, the LJ system has a higher shoulder for both $q^{d}(t)$ and $q^{l}(t)$, which indicates its dynamics is slower than the corresponding WCA system. This is also consistent with our previous results. More importantly, the dense-packed particles relax more slowly than the loose-packed particles by comparing $q^{d}(t)$ and $q^{l}(t)$ of the LJ system and WCA system. This phenomenon is more pronounced at lower temperatures (data of $\rho=1.2$ for several temperatures for both LJ and WCA systems is shown in Fig. S3 of the Supplemental Material~\cite{SM}). This is reasonable since dense-packed particles may have less free volume to relax than the loose-packed particles. This also suggests that there exists direct link between local density and dynamic heterogeneity for glass-forming liquids. The relatively dense-packed particles relax slowly and dominant the slow relaxation dynamics.

Our results indicate that the transiently localized cages, which trap the dynamics of particles, may have a length scale much larger than the size of fist coordination shell, and growing in its size as further supercooled. The relatively dense-packed structural environment may be the structural presentation of low potential energy domains, since it has been found that particles with low potential energy move slowly and clusters formed by them correlate with the dynamic heterogeneity~\cite{Matharoo}. These findings support that the liquid-glass transition is not solely a dynamic behavior but also accompanied with subtle changes of structures.

In Summary, we present evidences of the intrinsic link between the microscopic local density and the slow dynamics as well as dynamic heterogeneity of the glass-forming liquids. We find, at a large density range, that the difference of the dynamics of the LJ and WCA systems can be completely captured by the cluster size characterized by the relatively dense-packed particles, although the normally used static pair correlations and the statistical property of the Voronoi volume remain blind to the dynamics. The extracted static length scale of the dense-packed particles directly corresponds to the degree of supercooling and also the relaxation time of the glass-formers. The temporarily but continuously formed locally dense-packed structural environment may be the origin of the slow dynamics and dynamic heterogeneity in the glass-forming liquids. Our findings open a new way to study the central problem of glass transition and may also stimulate more experimental work on the structural nature of slow dynamics and dynamic heterogeneity.

This work is subsidized by the National Basic Research Program of China (973 Program, 2012CB821500), and supported by the National Natural Science Foundation of China (21222407, 21474111) program.


\begin{thebibliography}{9}

\bibitem{Berthier0}L.~Berthier and G.~Biroli, Rev. Mod. Phys. {\bf 83}, 587 (2011).

\bibitem{Binderbook}K.~Binder and W.~Kob, \emph{Glassy Materials and Disordered Solids} (World Scientific, Singapore, 2005).

\bibitem{Barrat}L.~Berthier, G.~Biroli, J.~-P.~Bouchaud, L.~Cipelletti, and W.~van~Saarloos, Dynamical Heterogeneities in Glasses, Colloids, and Granular Media (Oxford University Press, Oxford, 2011).

\bibitem{Karmakar0}S.~Karmakar, C.~Dasgupta, and S.~Sastry, Annu. Rev. Condens. Matter Phys. {\bf 5}, 255 (2014).

\bibitem{Berthier1}L.~Berthier and G.~Tarjus, Phys. Rev. Lett. {\bf 103}, 170601 (2009).

\bibitem{Berthier2}L.~Berthier and G.~Tarjus, J. Chem. Phys. {\bf 134}, 214503 (2011).

\bibitem{Berthier3}L.~Berthier and G.~Tarjus, Phys. Rev. E {\bf 82}, 031502 (2010).

\bibitem{Tanaka0}H.~Tanaka, T.~Kawasaki, H.~Shintani, and K.Watanabe, Nat. Mater. {\bf 9}, 324 (2010).

\bibitem{Tanaka1}K.~Watanabe, T.~Kawasaki, and H.~Tanaka, Nat. Mater. {\bf 10}, 512 (2011).

\bibitem{Tanaka2}T.~Kawasaki, T.~Araki, and H.~Tanaka, Phys. Rev. Lett. {\bf 99}, 215701 (2007).

\bibitem{Coslovich0}D.~Coslovich, Phys. Rev. E {\bf 83}, 051505 (2011).

\bibitem{Hocky1}G.~M.~Hocky, D.~Coslovich, A.~Ikeda and D.~R.~Reichman, Phys. Rev. Lett. {\bf 113}, 157801 (2014).

\bibitem{Bruning}R.~Br\"{u}ning, D.~A.~St-Onge, S.~Patterson, and W.~Kob, J. Phys.: Condens. Matter {\bf 21}, 035117 (2009).

\bibitem{Charbonneau}B.~Charbonneau, P.~Charbonneau, and G.~Tarjus, Phys. Rev. Lett. {\bf 108}, 035701 (2012).

\bibitem{Biroli0}J.~-P.~Bouchaud and G.~Biroli, J. Chem. Phys. {\bf 121}, 7347 (2004).

\bibitem{Biroli1}G.~Biroli, J.~-P.~Bouchaud, A.~Cavagna, T.~S.~Grigera and P.~Verrocchio, Nat. Phys. {\bf 4}, 771 (2008).

\bibitem{Kob1}W.~Kob, S.~Roldan-Vargas and L.~Berthier, Nat. Phys. {\bf 8}, 164 (2012).

\bibitem{Hocky}G.~M.~Hocky, T.~E.~Markland and D.~R.~Reichman, Phys. Rev. Lett. {\bf 108}, 225506 (2012).

\bibitem{Tarjus1}P.~Charbonneau and G.~Tarjus, Phys. Rev. E {\bf 87}, 042305 (2013).

\bibitem{Li}Y.~W.~Li, W.~S.~Xu, and Z.~Y.~Sun, J. Chem. Phys. {\bf 140}, 124502 (2014).

\bibitem{Tanakaadd}J.~Russo and H.~Tanaka, Proc. Natl. Acad. Sci. U.S.A. {\bf 112}, 6920 (2015).

\bibitem{cage1}L.~Larini, A.~Ottochian, C.~De~Michele, and D.~Leporini, Nat. Phys. {\bf 4}, 42 (2008).

\bibitem{cage2}C.~De~Michele, E.~Del~Gado, and D.~Leporini, Soft Matter {\bf 7}, 4025 (2011).

\bibitem{cage3}P.~Charbonneau, Y.~Jin, G.~Parisi, and F.~Zamponi, Proc. Natl. Acad. Sci. U.S.A. {\bf 111}, 15025 (2014).

\bibitem{cage4}P.~Charbonneau, A.~Ikeda, G.~Parisi, and F.~Zamponi, Proc. Natl. Acad. Sci. U.S.A. {\bf 109}, 13939 (2012).

\bibitem{Bernini1}S.~Bernini, F.~Puosi, and D.~Leporini, J. Non-Cryst. Solids {\bf 407}, 29 (2015).

\bibitem{Bernini2}S.~Bernini, F.~Puosi, and D.~Leporini, J. Chem. Phys. {\bf 142}, 124504 (2015).

\bibitem{Kob2}W.~Kob and H.~C.~Andersen, Phys. Rev. Lett. {\bf 73}, 1376 (1994).

\bibitem{WCA}J.~D.~Weeks, D.~Chandler, and H.~C.~Andersen, J. Chem. Phys. {\bf 54}, 5237 (1971).

\bibitem{SM}See Supplemental Material at http://... for some sytem information, results obtained from another definition of densely packed particles, discussions on some fitting parameters and test of the generality of our method, which includes Refs. ~\cite{sNose,sHoover,szhu,sWahn,sKim,sKob} for more details.

\bibitem{sNose}S.~Nos\'{e}, Mol. Phys. {\bf 52}, 255 (1984).

\bibitem{sHoover}W.~G.~Hoover, Phys. Rev. A {\bf 31}, 1695 (1985).

\bibitem{szhu}Y.~L.~Zhu, H.~Liu, Z.~W.~Li, H.~J.~Qian, G.~Milano, and Z.~Y.~Lu, J. Comput. Chem. {\bf 34}, 2197 (2013).

\bibitem{sWahn}G.~Wahnstr\"{o}m, Phys. Rev. A {\bf 44}, 3752 (1991).

\bibitem{sKim}K.~Kim and S.~Saito, J. Chem. Phys. {\bf 138}, 12A506 (2013).

\bibitem{sKob}W.~Kob, L.~Berthier, Phys. Rev. Lett. {\bf 110}, 245702 (2013).

\bibitem{xuning}L.~Wang and N.~Xu, Phys. Rev. Lett. {\bf 112}, 055701 (2014).

\bibitem{voronoi}J.~L.~Finney, Proc. R. Soc. London, Ser. A {\bf 319}, 479 (1970).

\bibitem{Starr0}F.~W.~Starr, S.~Sastry, J.~F.~Douglas, S.~C.~Glotzer, Phys. Rev. Lett. {\bf 89}, 125501 (2002).

\bibitem{Cheng}X.~Cheng, Soft Matter {\bf 6}, 2931 (2010).

\bibitem{Starr1}J.~C.~Conrad, F.~W.~Starr, and D.~A.~Weitz, J. Phys. Chem. B {\bf 109}, 21235 (2005).

\bibitem{xu}W.~S.~Xu, Z.~Y.~Sun, and L.~J.~An, Phys. Rev. E {\bf 86}, 041506 (2012).

\bibitem{Russo}J.~Russo and H.~Tanaka, Sci. Rep. {\bf 2}, 505 (2012).

\bibitem{Tanpeng}P.~Tan, N.~Xu, and L.~Xu, Nat. Phys. {\bf 10}, 73 (2014).

\bibitem{Matharoo}G.~S.~Matharoo, M.~S.~Gulam~Razul, and P.~H.~Poole, Phys. Rev. E {\bf 74}, 050502(R) (2006).




\end{thebibliography}
\end{document}


\title{Supplemental Material

for

Structural Signature of Slow and Heterogeneous Dynamics in Glass-Forming Liquids}
\author{Yan-Wei Li}
\affiliation{State Key Laboratory of Polymer Physics and Chemistry, Changchun Institute of Applied Chemistry, Chinese Academy of Sciences, Changchun 130022, China}
\author{You-Liang Zhu}
\affiliation{State Key Laboratory of Polymer Physics and Chemistry, Changchun Institute of Applied Chemistry, Chinese Academy of Sciences, Changchun 130022, China}
\author{Zhao-Yan Sun}
\email{zysun@ciac.ac.cn}
\affiliation{State Key Laboratory of Polymer Physics and Chemistry, Changchun Institute of Applied Chemistry, Chinese Academy of Sciences, Changchun 130022, China}
\maketitle

\subsection{The system information}

\begin{figure}[b]
 \centering
 \includegraphics[angle=0,width=0.48\textwidth]{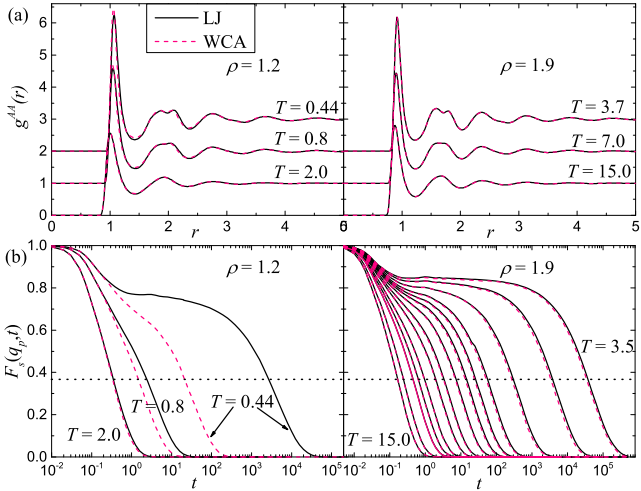}
 \renewcommand{\figurename}{FIG. S}
\caption{(a) The radial distribution functions (results have been shifted for clarity) and (b) the self-intermediate scattering functions at different temperatures for $\rho=1.2$ (left) and $\rho=1.9$ (right) for both LJ (black solid lines) and WCA (pink dashed lines) systems. The horizontal dotted line in (b) marks $F_{s}(q_{p},t)=1/e$ at which the relaxation time $\tau_{\alpha}$ is determined. All the results are for $A$ particles.}
\end{figure}

The present studied systems consist of $N=3074-4865$ (depending on $\rho$) with $80\%$ $A$ and $20\%$ $B$ particles. The potential between particles $\alpha$ and $\beta$ is $V_{\alpha\beta}(r) = 4\epsilon_{\alpha\beta}[(\sigma_{\alpha\beta}/r_{\alpha\beta})^{12} - (\sigma_{\alpha\beta}/r_{\alpha\beta})^{6} + C_{\alpha\beta}]$, when $r_{\alpha\beta}$ is smaller than the potential cutoff $r^{c}_{\alpha\beta}$, and zero otherwise. Here, $\alpha, \beta \in$ $\{A, B\}$. The interaction parameters are given by $\sigma_{AB}/\sigma_{AA}=0.8$, $\sigma_{BB}/\sigma_{AA}=0.88$, $\epsilon_{AB}/\epsilon_{AA}=1.5$, and $\epsilon_{BB}/\epsilon_{AA}=0.5$. The potential is truncated and shifted at $r^{c}_{\alpha\beta}$, i.e., $2^{1/6}\sigma_{\alpha\beta}$ for WCA system and $2.5\sigma_{\alpha\beta}$ for LJ system.  $C_{\alpha\beta}$ guarantees $V_{\alpha\beta}(r^{c}_{\alpha\beta})=0$. Length, energy and time are recorded in units of $\sigma_{AA}$, $\epsilon_{AA}$ and $\sqrt{m\sigma_{AA}^{2}/\epsilon_{AA}}$, respectively. The simulations are performed in the $NVT$ ensemble with periodic boundary conditions. The Newton's equations of motion were integrated with the velocity form of the Verlet algorithm and the temperature $T$ was maintained by the Nos\'{e}-Hoover thermostat~\cite{Nose,Hoover}. In order to obtain reliable results and improve the statistics, for each state point, we average the results over eight independent runs. All the simulations are performed with in-house GALAMOST software package~\cite{zhu}.

\begin{figure}[tb]
 \centering
 \includegraphics[angle=0,width=0.45\textwidth]{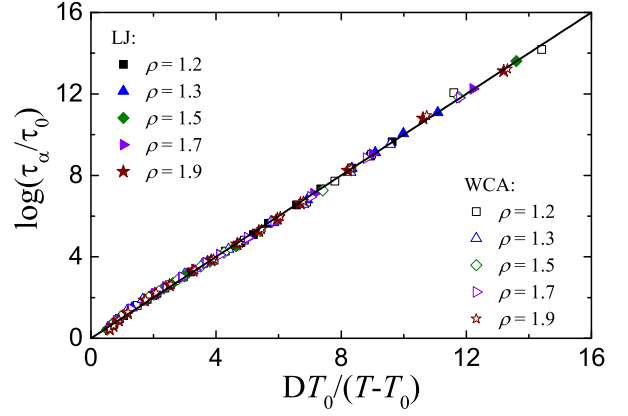}
 \renewcommand{\figurename}{FIG. S}
\caption{$T$ dependence of the relaxation time $\tau_{\alpha}$. Data are shown in the VFT form $log(\tau_{\alpha}/\tau_{0}) \sim DT_{0}/(T-T_{0})$, for all the temperatures at different densities of the LJ and WCA systems. The black solid line is the relation of $log(\tau_{\alpha}/\tau_{0})=DT_{0}/(T-T_{0})$.}
\end{figure}

We measure the intermediate scattering functions of $A$ particles, $F_{s}(q_{p},t)=\frac{1}{N_{A}}<\sum_{j=1}^{N_{A}}\exp\{i\textbf{q}_{p}\cdot[\textbf{r}_{j}(t)-\textbf{r}_{j}(0)]\}>$, where $\textbf{r}_{j}$ is the position vector of particle $j$, $<\cdot\cdot\cdot>$ indicates the thermal average, $i=\sqrt{-1}$ and the wave number $q_{p} = 7.3$ corresponds to the first peak of the static structure factor. The similarity of the static pair correlations and differences of the relaxation dynamics at different degrees of supercooling for $\rho=1.2$ and $\rho=1.9$ are shown in Fig. S1. Clearly, the pair correlations cannot account for the differences in dynamics for glass-forming liquids.

We show in Fig. S2 the VFT (Eq. (3) in the paper) fitting results of temperature dependence of the relaxation time in the whole density range we studied for both LJ and WCA systems. Clearly, the $T$ dependence of $\tau_{\alpha}$ remarkably agree with the VFT equation for all the state points investigated.

\begin{figure}[htb]
 \centering
 \includegraphics[angle=0,width=0.45\textwidth]{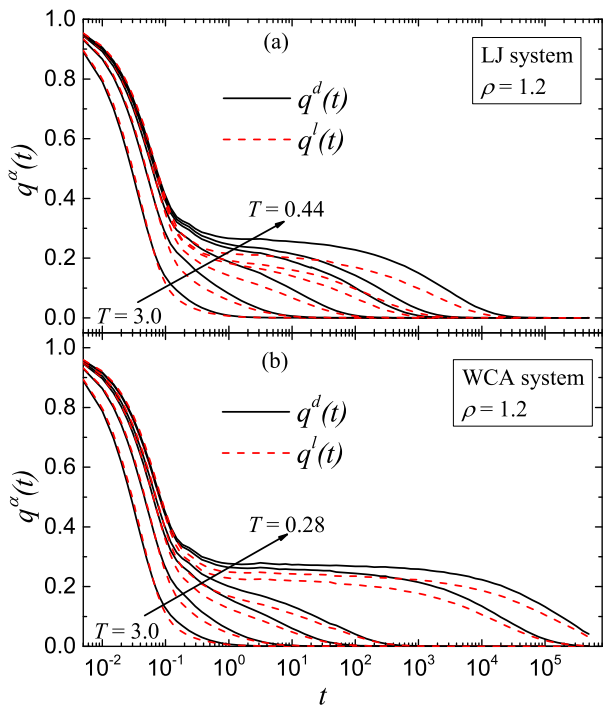}
 \renewcommand{\figurename}{FIG. S}
\caption{Time dependence of the self part of densely packed particles overlap $q^{d}(t)$ (black solid lines) and that of loosely packed particles overlap $q^{l}(t)$ (red dashed lines) for several temperatures at $\rho=1.2$ for both the LJ system (a) and the WCA system (b). Temperatures for the LJ system are $3.0,1.0,0.6,0.5,0.48$ and $0.44$ from left to right in (a). Temperatures for the WCA system are $3.0,1.0,0.5,0.4,0.3$ and $0.28$ from left to right in (b).}
\end{figure}

The comparison of $q^{d}(t)$ and $q^{l}(t)$ at different temperatures for $\rho=1.2$ in both the LJ system and the WCA system can be seen in Fig. S3. The curves show one-step decay at higher temperatures and two-step decay at lower temperatures, which is akin to the $T$ dependence of the intermediate scattering functions. The $q^{d}(t)$ curves always lie up the curves of $q^{l}(t)$ before decaying to $0$ even at high temperatures. This shows that the locally dense-packed particles always relax more slowly than the loosely packed particles, and this phenomenon is more pronounced in the supercooled regime. Results for the high $\rho$ data show similar behavior (data not shown). Thus, it can be concluded that the dynamic heterogeneity has a clear structural origin, i.e., the heterogeneity of local density.

\subsection{Results obtained from another definition of densely packed particles}

\begin{figure}[htb]
 \centering
 \includegraphics[angle=0,width=0.48\textwidth]{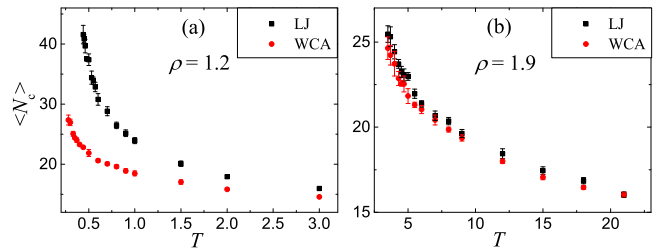}
 \renewcommand{\figurename}{FIG. S}
\caption{$T$ dependence of the number averaged cluster size $<N_{c}>$ of the LJ and WCA systems for $\rho=1.2$ (a) and $\rho=1.9$ (b). $<N_{c}>$ is obtained by averaging over 1000 configurations in each run, and the vertical error bars are obtained by averaging over 8 independent runs. Here the definition of densely packed particles is different from that in the main text (see SM text).}
\end{figure}

To study whether the definition of dense-packed particles influences our results, we check our data by using another definition of them. As described in the main text, the statical distribution of the normalized Voronoi volumes is universal in the density range we studied for both LJ and WCA systems. Inspired by it, particles whose Voronoi volume satisfies $(v-<v>)/\sigma_{v}<-1.2$, are defined as the dense-packed ones. Based on this definition, we calculate the averaged cluster size $<N_{c}>$ and show in Fig. S4 the $<N_{c}>$ data of $\rho=1.2$ and $\rho=1.9$ for both the LJ and WCA systems. Compared with results shown in Fig. 2 in the main text, the values of $<N_{c}>$ are smaller since the number of densely packed particles approximately accounts for $10\%$ of the total particles and this percentage is less than $13\%$ defined in the main text. Despite of quantitative differences, the qualitative results are very similar for both these two definitions of dense-packed particles. At $\rho=1.2$, $<N_{c}>$ is larger for LJ system and curves show big difference from WCA system. However, at $\rho=1.9$ the two curves are very close, which implies that $<N_{c}>$ has a clear connection with the dynamics, as has been discussed in the main text.

\begin{figure}[tb]
 \centering
 \includegraphics[angle=0,width=0.45\textwidth]{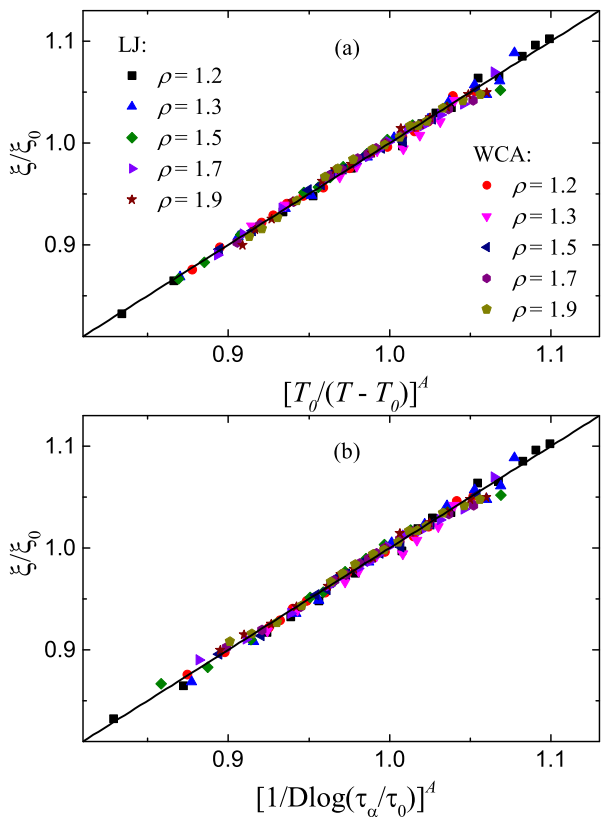}
 \renewcommand{\figurename}{FIG. S}
\caption{(a) The rescaled length scale $\xi/\xi_{0}$ versus $[T_{0}/(T-T_{0})]^{A}$. The black solid line is the relation of $\xi/\xi_{0}=[T_{0}/(T-T_{0})]^{A}$. (b) Relation between $\xi/\xi_{0}$ and $[1/Dlog(\tau_{\alpha}/\tau_{0})]^{A}$. The symbols are the same as in (a). The black solid line is the relation of $\xi/\xi_{0}=[1/Dlog(\tau_{\alpha}/\tau_{\alpha})]^{A}$. The definition of densely packed particles is shown in the SM text.}
\end{figure}

We extract the static length scale from the above $<N_{c}>$ data by using the function $\xi=\sqrt[3]{<N_{c}>}$, and we show in Fig. S5 the fitting results of Eq. (2) and Eq. (4) of the main text. Similar to Fig. 3(b) and 3(c) of the main text, data collapse on the line defined by these two equations, indicating that the defined static length scale correlates well with the degree of supercooling and also the relaxation time. All the above results indicate that the definition of the densely packed particles has no qualitative influence on the results, and a reasonable definition is expected to obtain similar qualitative results.

\subsection{Discussions on some fitting parameters}

\begin{figure}[htb]
 \centering
 \includegraphics[angle=0,width=0.35\textwidth]{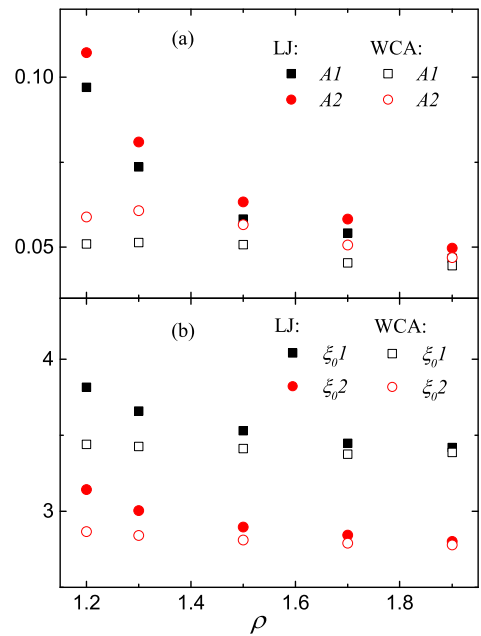}
 \renewcommand{\figurename}{FIG. S}
\caption{Density $\rho$ dependence of the fitting parameter $A$ (a) and $\xi_{0}$ (b) for both LJ and WCA systems. $A1$ and $\xi_{0}1$ are obtained based on the definition of densely packed particles used in the main text while $A2$ and $\xi_{0}2$ are obtained based on the definition of densely packed particles used in the SM.}
\end{figure}

Next we discuss the values of $A$ and $\xi_{0}$ extracted from fitting the $T$ dependence of $\xi$ with a function of Eq. (2) shown in the main text. We plot in Fig. S6 their dependence on density $\rho$ for both LJ and WCA systems. Since the values quantitatively depend on the definition of the dense-packed particles, we obtain two sets of fitting parameters, i.e., $A1$ ($\xi_{0}1$) and $A2$ ($\xi_{0}2$), which correspond to the definition of densely packed particles used in the main text and in the SM, respectively. In Ref.~\cite{Tanaka0}, the temperature $T$ dependence of length scale ($\xi'$) extracted from the structural order parameter can be described by $\xi'=\xi'_{0}[T_{0}/(T-T_{0})]^{2/d}$, where $d$ is the spatial dimension of the system. Note that the adjustable parameter $\xi'_{0}$ used in their work is the order of the particle radius, but the fitting is poor for us to use this value as our $\xi_{0}$ in Eq. (2) of the main text. As a result, it is meaningless to compare the value of $A$ with $2/d$ reported in Ref.~\cite{Tanaka0}. We have checked that our $\xi_{0}$ value roughly corresponds to the static length scale at the onset temperature (or a little below than the onset temperature) for each density in both LJ and WCA systems. It has been reported that the temperature dependence of the static point-to-set length scale of the LJ and WCA systems collapse when the length is scaled by its onset temperature value and $T$ is scaled by the onset temperature~\cite{Hocky}, although in a relatively small temperature range. In a larger range of supercooling this scaling collapse is found to break down in the hard spheres system~\cite{Tarjus1}. Our results may provide a more universal relation between the static length scale and temperature, as shown in Eq. (2) in the main text. 

It can be seen from Fig. S6 that the $\rho$ dependence of these two sets of $A$ and $\xi_{0}$ are qualitatively similar. The value of $A$ (both $A1$ and $A2$) decreases as increasing $\rho$ for LJ system while this value has little dependence on $\rho$ for WCA system. The difference of value $A1$ (or $A2$) for the two systems shrinks as increasing $\rho$ and is about to disappear at $\rho=1.9$. Similar behavior occurs in $\xi_{0}1$ (or $\xi_{0}2$). This is quite reminiscent of the different dynamics for LJ and WCA systems at different $\rho$. Our results indicate that the two fitting parameters $A$ and $\xi_{0}$ are depending on the intrinsic properties of the system and are quantitatively impacted by the definition of densely packed particles. It should be noted that the range of the fitting parameters for different densities for both LJ and WCA systems is relatively small. However, we find that the fitting parameters are very sensitive to the fitting results, and a little change of the value of fitting parameters can lead to obvious deviations from our data. Thus, the data collapse of Eq. (2) in the main text is not due to certain small values of $A$ but due to the intrinsic system properties.

\subsection{Test of the generality of our method in other glass models}

To test whether it is general that the locally dense-packed structural environment is the structural signature of slow dynamics, we study other two different models. The first one is the Wahnstr\"{o}m (Wahn) system~\cite{Wahn}. It is a $50$$:$$50$ binary Lennard-Jones mixture of spheres whose interaction parameters are very different from the Kob-Andersen model. The interaction parameters for the Wahn model are $\sigma_{AB}/\sigma_{AA}=0.916$, $\sigma_{BB}/\sigma_{AA}=0.833$, $\epsilon_{AB}=\epsilon_{BB}=\epsilon_{AA}$, and $m_{B}/m_{A}=0.5$. The potentials are cut and shifted at $2.5\sigma_{\alpha\beta}$. In the following, simulation results will be described in terms of the reduced units $\sigma_{AA}$, $\epsilon_{AA}$ and $\sqrt{m\sigma_{AA}^{2}/\epsilon_{AA}}$ for length, temperature, and time, respectively. We investigate the Wahn system with $N_{W}=3000$ particles at the number density $\rho_{W}=1.297$ in the temperature range $T=5.0$ to $T=0.58$. The mode-coupling temperature for this system is $T_{c}\approx0.56$~\cite{Kim,Hocky1}.

The second system we investigate is a $50$$:$$50$ binary mixture of $N_{H}=3000$ harmonic spheres (Harm) with the number density $\rho_{H}=0.675$~\cite{Kob}. All particles have the same mass $m$ and the interaction potential is given by $V_{\alpha\beta}(r) = 0.5\epsilon(1-r/\sigma_{\alpha\beta})^{2}$, if $r_{\alpha\beta}<\sigma_{\alpha\beta}$ and $0$ otherwise. The interaction parameters are $\sigma_{AB}/\sigma_{AA}=1.2$, $\sigma_{BB}/\sigma_{AA}=1.4$. Length, energy and time are reported in units of $\sigma_{AA}$, $10^{-4}\epsilon$ and $\sqrt{m\sigma_{AA}^{2}/\epsilon}$, respectively. The temperature range we studied for this system is $T=50$ to $T=5.5$ and the mode-coupling temperature is $T_{c}\approx5.2$~\cite{Hocky1,Kob,Kob1}.

\begin{figure}[htb]
 \centering
 \includegraphics[angle=0,width=0.48\textwidth]{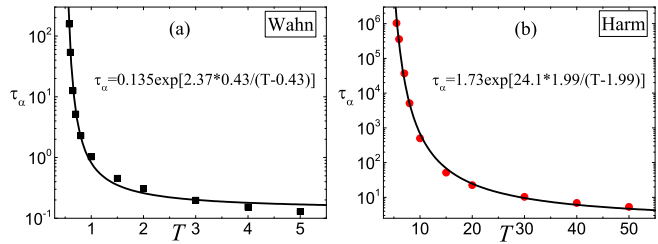}
 \renewcommand{\figurename}{FIG. S}
\caption{$T$ dependence of the $\alpha$ relaxation time $\tau_{\alpha}$ of the Wahn system (a) and the Harm system (b). Solid lines are the fitting results by the VFT relation (see Eq. (3) in the main text). Here $\tau_{\alpha}$ is obtained from $F_{s}(q_{p},\tau_{\alpha})=1/e$ of the $A$ part particles. The wave vector $q_{p}$ is equal to $7.7$ and $6.28$ for the Wahn and Harm system, respectively.}
\end{figure}

We also perform the molecular dynamics simulation with in-house GALAMOST software package~\cite{zhu} for the above two systems. Temperature is maintained by the Nos\'{e}-Hoover thermostat and the integration time steps are 0.005 and 0.04 for the Warn system and the Harm system, respectively. Results are also averaged over eight independent runs for the two systems.

We present the $T$ dependence of $\tau_{\alpha}$ and also the VFT fitting results for the above two systems in Fig. S7. Clearly, the VFT relation provides a good description of the results for both the Wahn and the Harm systems, especially at lower temperatures.

\begin{figure}[htb]
 \centering
 \includegraphics[angle=0,width=0.48\textwidth]{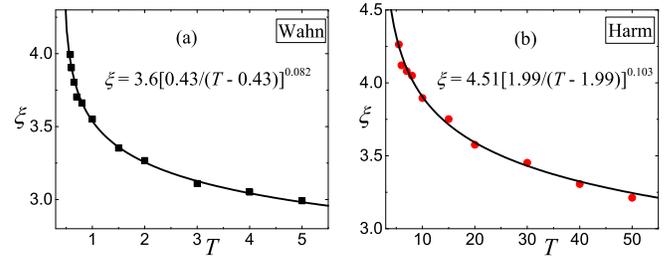}
 \renewcommand{\figurename}{FIG. S}
\caption{$T$ dependence of the length scale $\xi$ of the Wahn system (a) and the Harm system (b). Solid lines are the fitting results by Eq. (2) in the main text.}
\end{figure}

\begin{figure}[htb]
 \centering
 \includegraphics[angle=0,width=0.45\textwidth]{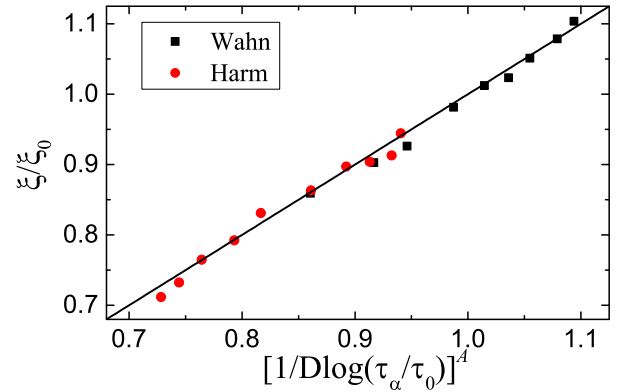}
 \renewcommand{\figurename}{FIG. S}
\caption{Relation between $\xi/\xi_{0}$ and $[1/Dlog(\tau_{\alpha}/\tau_{0})]^{A}$ for the Wahn system (black squares) and the Harm system (red circles). The black solid line is the relation of $\xi/\xi_{0}=[1/Dlog(\tau_{\alpha}/\tau_{\alpha})]^{A}$. Since the deviation from the VFT relation (see Fig. S8) for the data points of the two highest temperatures of the Wahn system, we didn't provide the data of $T=5.0$ and $T=4.0$ in this figure.}
\end{figure}

To obtain the static length scales of these two systems, we also define $13\%$ of particles with the largest local density as the locally dense-packed particles (this definition is the same as that in the main text). Then, we evaluate the number averaged cluster size $<N_{c}>$ formed by the densely packed particles. The static length scale is thus obtained from $\xi=\sqrt[3]{<N_{c}>}$. We show in Fig. S8 the temperature dependence of the length scale $\xi$ for both Wahn and Harm systems. Clearly, $\xi$ increases as decreasing temperature for both these two systems, which indicates that this kind of definition of $\xi$ can also capture the static structure differences between high temperature and low temperature glass-forming liquids. On the other hand, the $T$ dependence of $\xi$ can be well fitted by Eq. (2) of the main text, indicating that the static length scale has a correspondence relationship with the degree of supercooling (characterized by $T_{0}/(T-T_{0})$) for both the Warn and the Harm systems.

From Fig. S7 and Fig. S8, we can easily obtain the relationship between $\xi$ and $\tau_{\alpha}$ (see Eq. (4) in the main text). In Fig. S9, we plot $\xi/\xi_{0}$ against $[1/Dlog(\tau_{\alpha}/\tau_{0})]^{A}$. Slight deviations from perfect data collapse can be seen, which is because of the narrow range of the scaled $\xi$ and the scaled $\tau_{\alpha}$, and little deviation of the data points from the fitting line in Fig. S7 or Fig. S8 can lead to obvious deviations in Fig. S9. Nevertheless, the correlation between $\xi$ and relaxation time is striking. Thus, we conclude that the locally dense-packed structural environment can also be an important structural signature of dynamics for both Wahn and Harm systems.